\documentclass[aoas,MSNbibl,nameyear,dvips]{arximspdf}
\usepackage{multirow,dcolumn}

\doi{10.1214/13-AOAS709} 
\volume{8}
\issue{1}
\pubyear{2014}
\firstpage{595}
\lastpage{621}

\makeatletter
\newcolumntype{d}[1]{D{.}{.}{#1}}
\newtheorem{theorem}{Theorem}%
\newtheorem{cor}[theorem]{Corollary}%
\newtheorem{prop}[theorem]{Proposition}%
\newtheorem{lemma}[theorem]{Lemma}%
\newproclaim{condition}{Condition}%
\newproclaim{assumption}{Assumption}%
\newproclaim{example}{Example}%
\newproclaim{remark}{Remark}%
\newproclaim{definition}{Definition}%

\newcommand{\Z}{\mathbb{Z}}%
\newcommand{\R}{\mathbb{R}}%
\newcommand{\Aa}{\mathcal{A}}%
\newcommand{\Ll}{\mathcal{L}}%
\newcommand{\B}{\mathcal{B}}%
\newcommand{\F}{\mathcal{F}}%
\newcommand{\cG}{\mathcal{G}}%
\newcommand{\cJ}{\mathcal{J}}%
\newcommand{\cH}{\mathcal{H}}%
\newcommand{\bR}{\mathbf{R}}
\newcommand{\bw}{\mathbf{w}}%
\newcommand{\bu}{\mathbf{u}}%

\newcommand{\bS}{\mathbf{S}}%
\newcommand{\bepsilon}{\bolds{\varepsilon}}%
\newcommand{\bo}{\mathbf{0}}%
\makeatother

\begin{document}
\begin{frontmatter}

\title{The role of the information set for forecasting---with
applications to risk management}
\runtitle{The role of the information set for forecasting}

\begin{aug}
\author[a]{\fnms{Hajo} \snm{Holzmann}\corref{}\ead[label=e1]{holzmann@mathematik.uni-marburg.de}\thanksref{t1}}
\and
\author[a]{\fnms{Matthias} \snm{Eulert}\ead[label=e2]{eulert@mathematik.uni-marburg.de}}
\thankstext{t1}{Supported in part by the DFG, Grant Ho 3260/3-1.}
\address[a]{Fachbereich Mathematik und Informatik\\
Philipps-Universit{\"a}t Marburg\\
Hans-Meerwein-Stra{\ss}e\\
35032 Marburg\\
Germany\\
\printead{e1}\\
\phantom{E-mail:\ }\printead*{e2}}
\affiliation{Philipps-Universit{\"a}t Marburg}
\runauthor{H.~Holzmann and M.~Eulert}
\end{aug}

\received{\smonth{10} \syear{2012}}
\revised{\smonth{12} \syear{2013}}

%
\begin{abstract}
Predictions are issued on the basis of certain information. If the
forecasting mechanisms are correctly specified, a larger amount of
available information should lead to better forecasts. For point
forecasts, we show how the effect of increasing the information set can
be quantified by using strictly consistent scoring functions, where it
results in smaller average scores. Further, we show that the classical
Diebold--Mariano test, based on strictly consistent scoring functions
and asymptotically ideal forecasts, is a consistent test for the effect
of an increase in a sequence of information sets on $h$-step point
forecasts. For the value at risk (VaR), we show that the average score,
which corresponds to the average quantile risk, directly relates to the
expected shortfall. Thus, increasing the information set will result in
VaR forecasts which lead on average to smaller expected shortfalls. We
illustrate our results in simulations and applications to stock returns
for unconditional versus conditional risk management as well as
univariate modeling of portfolio returns versus multivariate modeling
of individual risk factors. The role of the information set for
evaluating probabilistic forecasts by using strictly proper scoring
rules is also discussed.
\end{abstract}
%

\end{frontmatter}

\section{Introduction}
Making and evaluating statistical forecasts is a basic task for
statisticians and econometricians. While probabilistic forecasts,
consisting of a complete predictive distribution, are most informative
[cf. \citet{MR2325275}], interest often focuses on single-value point
forecasts [\citet{MR2847988}]. For example, in quantitative risk
management, the goal is to estimate certain functionals of a predictive
distribution such as the value at risk (VaR) or the expected shortfall
[\citet{MR2175089}].

Forecasts are issued on the basis of certain information. Evidently,
increasing the information set should lead to better forecasts, at
least if the forecasting mechanisms are correctly specified. We shall
call such forecasts ideal. In this article, we show how an improvement
of ideal forecasts by increasing the information set can be quantified
by using strictly consistent scoring functions [\citet{MR2847988}],
where it results in smaller average scores. Further, we show that the
classical \citet{Diebold1995CPA} test, based on strictly consistent
scoring functions and asymptotically ideal forecasts, is a consistent
test for the effect of an increase in a sequence of information sets on
$h$-step point forecasts.

As a most important example, consider evaluating VaR forecasts.
Formally, the VaR is a (high, say, $0.99$ or $0.999$) quantile of the
loss distribution. Unconditional methods base the VaR on the
unconditional distribution of the risk factors, thus using a trivial
information set, while conditional methods refer to a conditional
distribution typically given the historical data; see \citet
{MR2175089}. For conditional methods, the information set may vary as
well: in a portfolio point of view it only includes the portfolio
returns, while a modeling of the individual risk factors involves a
larger information set.

Unconditional backtesting consists in checking whether the relative
frequency of exceedances of the VaR estimates corresponds to the level
of the VaR. This is, as the name suggests, satisfied by both
unconditional and conditional methods if correctly specified.
Conditional methods are accompanied in case of one-step ahead estimates
by checking whether exceedances of VaR forecasts occur independently
[the i.i.d. hypothesis, cf. \citet{Christoffersen1998EIF,MR2175089}].
However, independence of exceedance indicators alone does not
adequately take into account the size of the information set for the
conditional methods; see also
\citet{Berkowitz2011EVAR}.


We show that by evaluating (ideal) VaR forecasts by scoring functions,
one can distinguish between VaR forecasts arising from distinct
information sets. Interestingly, increasing the information set will
result in VaR forecasts which lead to smaller expected shortfalls,
unless an increase in the information set does not result in any change
in the VaR forecast.

The paper is organized as follows. The general methodology is developed
in Section~\ref{sec:methodol}. To illustrate, we start in Section~\ref{sec:regressionintro} with an example from regression analysis. We
recall the well-known fact that by including additional variables and
thus increasing the information set, the mean-squared prediction error
of the (population, i.e., ideal) mean regression function is reduced.
Then, turning to general expectile regression, we indicate that our
subsequent results imply that including additional variables will
reduce the mean asymmetric squared loss of the (ideal)
expectile-regression functions. In Section~\ref{sec:constscore} we
show how the effect of a larger information set for issuing a certain
point forecast can be quantified by using strictly consistent scoring functions.
Section~\ref{sec:distrproprules} is concerned with the same problem in
case of evaluating probabilistic forecasts by using proper scoring
rules. See also the note by \citet{Tsyplakov2011EDFA}, which comments
on the paper by \citet{MR2843116} which in turn is a critical comment
on \citet{MR2325275}.
In Section~\ref{sec:testinginfo} we investigate the properties of the
\citet{Diebold1995CPA} test in the situation of nested sequences of
information sets and asymptotically ideal forecasts.

Section~\ref{sec:var} contains a detailed discussion of methods to
evaluate VaR forecasts. We start by discussing applications of the VaR
such as risk controls for trading desks, VaR-based portfolio choice and
regulatory uses, as well as general strategies for issuing VaR
forecasts. In Section~\ref{sec:exceedind} we focus on exceedance
indicators which are the typical tool for backtesting VaR forecasts,
and in Section~\ref{sec:quantileloss} we turn to the quantile loss
(the strictly consistent scoring function for the VaR) and relate its
expected value to the expected shortfall.

In Section~\ref{sec:simappl} we conduct a simulation study and give
applications to series of stock-returns for value at risk estimation,
when comparing first unconditional versus conditional methods and
second univariate modeling on the basis of portfolio returns versus
multivariate modeling of the individual risk factors. Section~\ref{sec:conclusions} concludes, while technical proofs are deferred to an \hyperref[app]{Appendix}.

\section{Quantifying the role of the information set}\label{sec:methodol}
%
\subsection{An introductory example from regression analysis}\label
{sec:regressionintro}

To motivate the upcoming discussion, consider an example in a
regression framework. Suppose that a triple $(Y, X_1, X_2)$ of random
variables is observed, where $Y$ is the dependent variable with $E |Y|
< \infty$ and $X_1, X_2$ are explanatory random variables.

Consider the mean regression $g(x_1, x_2) = E(Y | X_1 = x_1, X_2 =
x_2)$ of $Y$ on $(X_1, X_2)$, as well as $f(x_1) = E(Y | X_1 = x_1)$ of
$Y$ on $X_1$ only. Given values $x_1, x_2$, in which sense is $g(x_1,
x_2)$ a more precise forecast than $f(x_1)$ for the conditional mean of
$Y$, or phrased otherwise, in which sense is the forecast improved if
the information set is increased from
$\F= \sigma(X_1)$ to $\cG= \sigma(X_1, X_2)$?

As is well known, if $EY^2 < \infty$, we have that $P$-almost surely
($P$-a.s.)
\begin{eqnarray*}
E \bigl(\bigl(Y - g(X_1,X_2)
\bigr)^2|X_1 \bigr) & = &E \bigl(Y^2|\F \bigr)
- E \bigl(\bigl(E(Y|\cG)\bigr)^2|\F \bigr)
\\
& \leq &E \bigl(Y^2|\F \bigr) - \bigl(E(Y|\F)\bigr)^2 = E
\bigl(\bigl(Y - f(X_1)\bigr)^2|X_1 \bigr)
\end{eqnarray*}
since by the conditional Jensen inequality, $ (E(Y|\F))^2 \leq E
((E(Y|\cG))^2|\F )$, and therefore also the unconditional squared
forecast error is reduced:
\[
E \bigl(\bigl(Y - g(X_1,X_2)\bigr)^2 \bigr)
\leq E \bigl(\bigl(Y - f(X_1)\bigr)^2 \bigr).
\]
\citet{MR2899176} discuss the special case of mean prediction and the
effect of an increased information set in a dynamic context.
%
%

Now, the natural question is whether analogous statements are true if
we move away from the simple mean regression, say, to an expectile
regression on the $\alpha$ expectile, $\alpha\neq1/2$, or even
consider the whole predictive distributions $\Ll(Y|\F)$ and $\Ll
(Y|\cG)$.

Recall that the $\alpha$ expectile $\tau_\alpha$ of a distribution
function $F$ on $\R$ with finite first moment is defined as the unique
solution in $\tau$ to
\[
\alpha\int_\tau^\infty(y-\tau) \,dF(y) = (1-\alpha)
\int^\tau _{-\infty} (\tau-y) \,dF(y).
\]
Let $g_\alpha(x_1, x_2)$ [resp., $f_\alpha(x_1)$] denote the $\alpha$
expectile of the conditional distribution function of $Y$ given $X_1 =
x_1, X_2 = x_2$ (resp., given $X_1 = x_1$). Our result below implies
that if $E Y^2 < \infty$, and if we replace the squared loss $(y-m)^2$
for the mean by the asymmetric squared loss $S_\alpha(y, \tau) =
|1_{\tau\geq y} - \alpha|  (y-\tau)^2$ for the $\alpha$ expectile,
then $P$-a.s.
\[
E \bigl(S_\alpha\bigl(Y, g_\alpha(X_1,
X_2)\bigr) | \F \bigr) \leq E \bigl(S_\alpha\bigl(Y,
f_\alpha(X_1)\bigr) | \F \bigr)
\]
as well as
\[
E \bigl(S_\alpha\bigl(Y, g_\alpha(X_1,
X_2)\bigr) \bigr) \leq E \bigl(S_\alpha \bigl(Y,
f_\alpha(X_1)\bigr) \bigr)
\]
with equality if and only if $g_\alpha(X_1, X_2) = f_\alpha(X_1)$.
This will be deduced by using the fact that the above loss functions
are strictly consistent for the functionals, as defined below.

\subsection{Functionals and scoring functions}\label{sec:constscore}
We start by recalling the concept of strictly consistent scoring
functions; see \citet{MR2847988}. Let
$\Theta$ be a class of distribution functions on a closed subset $D
\subset\R$, which we identify with their associated probability
distributions, and let
$T\dvtx \Theta\to\R$ be a (one-dimensional) statistical functional.
We let $\mathcal{B}(\Theta)$ denote the Borel $\sigma$-algebra on
$\Theta$ w.r.t. the topology of weak convergence of distribution
functions (or probability measures), and we let $\mathcal B$ denote the
ordinary Borel $\sigma$-algebra on $\R$. We shall call the functional
$T$ measurable if it is $\mathcal{B}(\Theta) -\mathcal{B} $-measurable.

A \textit{scoring function} is a measurable map
$S\dvtx \R\times D \to[0, \infty)$.
Then $S(x,y)$ is interpreted as the loss if forecast $x$ is issued and
$y$ materializes.
$S$ is consistent for the functional $T$ relative to the class $\Theta
$ if
\[
\mbox{for all } x \in\R, F \in\Theta\dvtx \qquad E_F \bigl(S\bigl(T(F),Y
\bigr) \bigr) \leq E_F \bigl(S(x,Y ) \bigr),
\]
where $Y$ is a random variable with distribution function $F$, and we
assume that the relevant expected values exist and are finite. Thus,
the true functional $T(F)$ minimizes the expected loss under $F$.
If
\[
E_F \bigl(S\bigl(T(F),Y\bigr) \bigr) = E_F
\bigl(S(x,Y) \bigr)\qquad \mbox{implies that } x = T(F),
\]
then $S$ is \textit{strictly consistent} for $T$. If the functional $T$
admits a strictly consistent scoring function, then it is called \textit{elicitable} (relative to the class $\Theta$).
For several functionals such as mean, quantiles and expectiles \citet
{MR2847988} characterizes all strictly consistent scoring functions
which additionally satisfy the following:
%
\begin{eqnarray}
\label{eq:propscore}
&&\mathrm{1.} \ S(x,y) \geq0 \mbox{ with equality if
and only if } x=y,
\nonumber
\\
&&\mathrm{2.} \ S(x,y) \mbox{ is continuous in } x \mbox{ for all }y \in D,
\\
\quad &&\parbox{12cm}{$\mathrm{3.}$\ the partial derivative $\partial_x S(x,y)$
 exists and is continuous in  $x$  for $x \neq y$.}\nonumber %
\end{eqnarray}
Note that for simplicity we do not consider set-valued functionals. Our
results could be extended to include these, but the formulations would
become more cumbersome. Thus, in case of quantiles, we assume that all
distributions functions in $\Theta$ are strictly increasing.

\citet{MR2847988} also points out that well-known functionals such as
variance or expected shortfall are not elicitable. \citet
{Heinrich2013TMFI} obtains a corresponding negative result for the mode
functional, despite the convexity of the level sets for the mode.
%

Now let us consider a forecasting situation. Forecasts are issued on
the basis of certain information. Let $(\Omega, \Aa, P)$ be a
probability space, let $\F\subset\Aa$ be a sub-$\sigma$-algebra of
$\Aa$ (the information set), and let $Y\dvtx \Omega\to\R$ be a random
variable. The aim is to predict a particular functional of the
conditional distribution of $Y$ given~$\F$.
%
\begin{theorem}\label{the:basic}
Let $F_{ Y|\F}(\omega, \cdot)$ be the conditional distribution
function of $Y$ given $\F$. Assume that for each $\omega\in\Omega$,
$F_{ Y|\F}(\omega, \cdot) \in\Theta$.
If $T \dvtx \Theta\to\R$ is measurable, then $T(F) = T ( F_{ Y|\F
}(\omega, \cdot) ) = \hat Y(\omega)$ is an $\F$-measurable r.v.
If $T$ is elicitable (over $\Theta$) and if $S$ is a strictly
consistent scoring function for $T$, then for any $\F$-measurable
r.v. $Z$, we get
%
\begin{equation}
\label{eq:ineqfirst} E \bigl(S(\hat Y, Y) | \F \bigr) (\omega) \leq E \bigl(S(Z, Y) | \F
\bigr) (\omega)\qquad \mbox{for }P\mbox{-a.e. } \omega\in\Omega
\end{equation}
as well as for the mean scores that
%
\begin{equation}
\label{eq:meanloss} E \bigl(S(\hat Y, Y) \bigr) \leq E \bigl(S(Z, Y) \bigr)
\end{equation}
with equality in (\ref{eq:ineqfirst}) or (\ref{eq:meanloss}) if and
only if $\hat Y = Z$, $P$-a.s.
\end{theorem}
Let us turn to the situation where forecasts can be issued on the basis
of two distinct information sets $\F\subset\cG\subset\Aa$.
Evidently, the larger information set should only yield better ideal
forecasts and, indeed, we have the following result.
%
\begin{cor}\label{cor:condineq}
Suppose that $\F\subset\cG\subset\Aa$ are increasing information
sets. Set
%
\begin{equation}
\label{eq:increasinfo} \hat Y_\F(\omega) = T \bigl( F_{ Y|\F}(\omega,
\cdot) \bigr), \qquad\hat Y_\cG(\omega) = T \bigl( F_{Y|\cG}(\omega,
\cdot) \bigr).
\end{equation}
Then
%
\begin{eqnarray}
\label{eq:ineqsecondincrease} 
E \bigl(S(\hat Y_\cG, Y)|\cG \bigr) & \leq&
E \bigl(S(\hat Y_\F, Y)|\cG \bigr), \qquad P\mbox{-a.s.},
\nonumber\\
E \bigl(S(\hat Y_\cG, Y)|\F \bigr) & \leq& E
\bigl(S(\hat Y_\F, Y)|\F \bigr), \qquad P\mbox{-a.s.},
\\
\nonumber
E \bigl(S(\hat Y_\cG, Y) \bigr) & \leq& E
\bigl(S(\hat Y_\F, Y) \bigr), 
\end{eqnarray}
with equality in any of the inequalities in (\ref
{eq:ineqsecondincrease}) if and only if $\hat Y_\F= \hat Y_\cG$, $P$-a.s.
\end{cor}
Thus, increasing the information set always leads to better ideal
forecasts in terms of the score, except if the smaller information set
already gives the same forecasts for the corresponding functional.

Finally, we point out that the equality $\hat Y_\F= \hat Y_\cG$,
$P$-a.s. does not imply that the conditional distributions are equal,
as the following example shows.
%
\begin{example}\label{ex:examplequant}
We give an example involving quantiles. For a strictly increasing,
continuous distribution function $F$ let $q_\alpha(F)$ denote the
$\alpha$ quantile, $\alpha\in(0,1)$, and let $q_\alpha$ be the
$\alpha$ quantile of the standard normal distribution $N(0,1)$. Fix
$\alpha\in(0,1)$, $\sigma>1$, and let $B, X_1, X_2$ be independent
random variables with $B \sim Ber(1/2)$, $X_1 \sim N(0,1)$, $X_2 \sim
N (q_\alpha(1-\sigma), \sigma^2  ) $, and set $Y = B X_1 +
(1-B) X_2$. If $\F= \{ \varnothing, \Omega\}$ is trivial and $\cG=
\sigma\{ B\}$, then the conditional distributions of $Y$ are
\begin{eqnarray*}
\Ll(Y | \F) &=& \tfrac{1}{2} N(0,1) + \tfrac{1}{2} N
\bigl(q_\alpha (1-\sigma),\sigma^2 \bigr),\\
 \Ll(Y | \cG) &=& B
N(0,1) + (1-B) N \bigl(q_\alpha(1-\sigma),\sigma^2 \bigr),
\end{eqnarray*}
and in both cases the conditional $\alpha$ quantile is constant and
equals $q_\alpha$.
\end{example}
Indeed, in order to evaluate the complete forecast distribution,
strictly proper scoring rules are needed, as discussed in the next section.
%
\subsection{Probabilistic forecasts and proper scoring rules}\label
{sec:distrproprules}
Let us briefly discuss general proper scoring rules; see \citet
{MR2345548} for a detailed exposition. Recall that we identify the
distribution functions $F \in\Theta$ with their associated
probability measures $\mu_F \in\Theta$. A measurable mapping
$\bS\dvtx  \Theta\times D \to\R$ is called a \textit{scoring rule}. It is
called \textit{proper} if for any $\mu\in\Theta$,
%
\begin{equation}
\label{eq:properscorerule} E_\mu \bigl(\bS(\mu, Y) \bigr) \leq E_\mu
\bigl(\bS(\nu, Y) \bigr) \qquad\mbox{for all }\nu\in\Theta,
\end{equation}
and \textit{strictly proper} if there is equality in (\ref
{eq:properscorerule}) if and only if $\mu=\nu$.
\citet{MR2847988} points out that a functional $T$ together with a
consistent scoring function $S$ induces the proper scoring rule $\bS
(\mu_F,y) = S (T(F),y  )$. However, even if $S$ is strictly
consistent, $\bS$ will not necessarily be strictly proper.

Let again $(\Omega, \Aa, P)$ be a probability space, and let $\F
\subset\Aa$ be a sub-$\sigma$-algebra of $\Aa$ (the information
set). A \textit{Markov kernel} $G_\F$ [from $(\Omega, \F)$ to $(\R, \B
)$] is a mapping
\[
G_\F\dvtx \Omega\times\B\to[0, 1],
\]
such that:
\begin{longlist}[1.]
\item[1.] for any $\omega\in\Omega$, $B \mapsto G_\F(\omega, B)$ ($B \in
\B)$ is a probability measure on $(\R, \B)$,

\item[2.] for any $B \in\B$, $\omega\mapsto G_\F(\omega, B)$ is $\F- \B
[0,1]$-measurable.
\end{longlist}

The {\sl(regular) conditional distribution} $\mu_{Y| \F}$ of $Y$
given $\F$ is a particular Markov kernel [from $(\Omega, \F)$ to
$(\R, \B)$] such that for all $B \in\B$,
\[
E (1_{Y \in B} |\F ) (\omega) = \mu_{Y| \F}(\omega, B) \qquad\mbox{for }P
\mbox{-a.e. } \omega\in\Omega.
\]
%
\begin{theorem}\label{the:characpropscorerule}
Let $\bS$ be a strictly proper scoring rule. Let $\mu_{Y|\F}(\omega
, \cdot)$ be the conditional distribution of $Y$ given $\F$. Assume
that for each $\omega$, $\mu_{Y|\F}(\omega, \cdot) \in\Theta$.
For any Markov kernel $G_\F$ [from $(\Omega, \F)$ to $(\R, \B)$]
for which $G_\F(\omega, \cdot) \in\Theta$ for all $\omega\in
\Omega$, the map $\omega\mapsto\bS (G_\F(\omega,\cdot),
Y(\omega)  )$ is a random variable and we have that
%
\begin{equation}
\label{eq:ineqfirstrule} E \bigl(\bS(\mu_{Y| \F}, Y) | \F \bigr) (\omega) \leq E
\bigl(\bS(G_\F, Y) | \F \bigr) (\omega)\qquad \mbox{for }P\mbox{-a.e. } \omega \in\Omega
\end{equation}
and
%
\begin{equation}
\label{eq:ineqsecondrule} E \bigl(\bS(\mu_{Y| \F}, Y) \bigr)\leq E \bigl(
\bS(G_\F, Y) \bigr)
\end{equation}
with equality in (\ref{eq:ineqfirstrule}) or (\ref
{eq:ineqsecondrule}) if and only if for $P\mbox{-a.e. } \omega\in\Omega$,
the distributions $G_\F(\omega, \cdot)$ and $\mu_{Y| \F}(\omega,
\cdot)$ coincide.
\end{theorem}
This is also observed in \citet{Tsyplakov2011EDFA} in his comment on
the paper by \citet{MR2843116} which in turn was a critical response
to \citet{MR2325275}. \citet{MR2325275} discuss the somewhat too
dominant role of the probability integral transform (PIT) in evaluating
forecasts. They focus on the uniformity of the PIT if the forecasts are
correctly specified. \citet{Tsyplakov2011EDFA} also indicates a result
similar to Proposition~\ref{prop:independ} (see Section~\ref{sec:exceedind}) for the PIT and observes that mere independence of the
PIT values does not adequately take into account the role of the
information set.
%
\begin{cor}\label{cor:scoreruleincreaseinfo}
Let $\F\subset\cG\subset\Aa$ be increasing information sets. If
$\bS$ is a strictly proper scoring rule and for each $\omega$, $\mu
_{Y|\F}(\omega, \cdot), \mu_{Y|\cG}(\omega, \cdot) \in\Theta$, then
%
\begin{equation}
\label{eq:increasrulefirst} E \bigl(\bS(\mu_{Y| \cG}, Y) | \cG \bigr) (\omega) \leq E
\bigl(\bS(\mu_{Y| \F}, Y)| \cG \bigr) (\omega) \qquad\mbox{for }P\mbox{-a.e. } \omega\in\Omega
\end{equation}
and
%
\begin{equation}
\label{eq:ineqsecondrule2} E \bigl(\bS(\mu_{Y| \cG}, Y) \bigr)\leq E \bigl(\bS(
\mu_{Y| \F
}, Y) \bigr),
\end{equation}
with equality in (\ref{eq:increasrulefirst}) or (\ref
{eq:ineqsecondrule2}) if and only if for $P\mbox{-a.e. } \omega\in\Omega$,
the conditional distributions $\mu_{Y| \cG}(\omega, \cdot)$ and
$\mu_{Y| \F}(\omega, \cdot)$ coincide.

If in particular $\cG= \sigma(\F, \cH)$, where $\cH\subset\Aa$
is another sub-$\sigma$-algebra, then there is equality in (\ref
{eq:increasrulefirst}) or (\ref{eq:ineqsecondrule2}) if and only if
$Y$ and $\cH$ are conditionally independent given $\F$.
\end{cor}
Thus, using a strictly proper scoring rule to evaluate the complete
predictive distribution, the predictive distributions in Example~\ref
{ex:examplequant} based on distinct information sets could be
distinguished. However, if interest is focused on a single functional
like the mean or the VaR, then this might not be necessary. The second
part of the corollary extends results by \citet{Broecker2009RSAT} and
\citet{Degroot1983TCAE} from finite to general real state space.
%


\subsection{Testing for sufficient information}\label{sec:testinginfo}
Consider the setting of Section~\ref{sec:constscore} in which the aim
is to forecast a functional $T\dvtx \Theta\to\R$.
When evaluating forecasts empirically, one observes a sequence of
forecasts $\hat Y_1, \ldots, \hat Y_N$ of $T$ with the corresponding
realizations $Y_1, \ldots, Y_N$ and proceeds by averaging the
corresponding scores.

More specifically, assume that $(Y_n)_{n \geq1}$ is a stationary and
ergodic sequence, and let $(\F_n)_{n \geq1}$ be a filtration
(increasing sequence of sub-$\sigma$-algebras of $\Aa$) such that
$Y_n$ is $\F_n$-measurable, $n \geq1$.
Suppose that the $h$-step forecasts
\[
\hat Y_{n, \F}^{(h)}(\omega): = \hat Y_{ \F_{n-h}}(\omega) =
T \bigl( F_{Y_n|\F_{n-h} }(\omega, \cdot) \bigr)
\]
are stationary and ergodic as well. Then for the averaged loss, as $N
\to\infty$,
%
\begin{equation}
\label{eq:converg} \widehat m_{N, \F}:= \frac{1}{N} \sum
_{n=1}^N S\bigl(\hat Y_{n, \F
}^{(h)},
Y_n\bigr) \to E \bigl(S\bigl(\hat Y_{1, \F}^{(h)},
Y_1\bigr) \bigr),\qquad P\mbox{-a.s.}
\end{equation}
In this section we investigate the behavior of the classical \citet
{Diebold1995CPA} test when evaluating asymptotically ideal forecasts
based on distinct, nested information sets using strictly consistent
scoring functions. See below for further discussion on the relation to
the literature.

Suppose that $(\cG_n)$ is a second filtration for which $\F_n \subset
\cG_n$ for all $n \geq1$, and for which the sequence $\hat Y_{n, \cG
}^{(h)}: = \hat Y_{\cG_{n-h}}$ is stationary and ergodic as well.
We shall propose a test for the hypothesis
%
\begin{equation}
\label{eq:nullhypothesis} H\dvtx \hat Y^{(h)}_{n, \cG} = \hat
Y^{(h)}_{n, \F},\qquad P\mbox{-a.s.} \mbox{ for all }n\geq1,
\end{equation}
that both sequences of information sets lead to the same forecasts.
By stationarity, this is equivalent to $\hat Y^{(h)}_{1, \cG} = \hat
Y^{(h)}_{1, \F}$, $P$-a.s.

The $h$-step forecasts for time $n$ based on $\cG_{n-h}$ and on $\F
_{n-h}$ which are actually issued are denoted by $\tilde Y_{n, \cG
}^{(h)} $ and $\tilde Y_{n, \F}^{(h)} $.
Since we are concerned with the ideal forecasts, we need to make the
rather strong assumption that the errors (due to misspecification and
estimation effects) in these sequences of forecasts have an
asymptotically negligible effect on the scores. More precisely,
consider the following conditions:
%
\begin{eqnarray}
\label{eq:asympnoeffect} %
\sum_{n=1}^N
\bigl(S\bigl(\tilde Y^{(h)}_{n,\cJ}, Y_n\bigr) - S
\bigl(\hat Y^{(h)}_{n,\cJ}, Y_n\bigr) \bigr) =
o_P(\sqrt{N})
\nonumber
\\[-8pt]
\\[-8pt]
\eqntext{\bigl(\mbox{or }  = O_P(\sqrt{N})
\bigr), \cJ= \F, \cG.} %
\end{eqnarray}
As a test statistic, consider
\[
M_N = \frac{1}{N} \sum_{n=1}^N
\bigl(S\bigl(\tilde Y^{(h)}_{n, \F}, Y_n\bigr) - S
\bigl(\tilde Y^{(h)}_{n,\cG}, Y_n\bigr) \bigr) =
\widehat m_{N, \F} - \widehat m_{N, \cG}.
\]
%
\begin{theorem}\label{the:testsuffinfo}
Under the above stationarity assumptions suppose that\break $E (S(\hat
Y^{(h)}_{1, \F}, Y_1)^2 ) < \infty$ is satisfied. Under the null
hypothesis $H$ in (\ref{eq:nullhypothesis}), if (\ref
{eq:asympnoeffect}) holds with $o_P(\sqrt{N})$, then
%
\begin{eqnarray}
\label{eq:asymptest}\qquad \sqrt{N} M_N & \stackrel{d} {\to} &N \bigl(0,
\sigma^2 \bigr),
\nonumber
\\[-8pt]
\\[-8pt]
\nonumber
\sigma^2 & =& E \Biggl(Z_1^2 + 2 \sum
_{n=2}^h Z_1 Z_n
\Biggr), \qquad Z_n = S \bigl(\hat Y^{(h)}_{n, \F},
Y_n \bigr) - S \bigl(\hat Y^{(h)}_{n,
\cG},
Y_n \bigr).
\end{eqnarray}
Under an alternative, if (\ref{eq:asympnoeffect}) holds with $O_P
(\sqrt{N} )$, we get $\sqrt{N} M_N \to\infty$ in probability.
\end{theorem}
Let us give some remarks on the above result.

1. Suppose that $\hat\sigma_N^2$ is a consistent estimate of the
long-run variance $\sigma^2$. Then form the $t$-statistic
\[
T_N = \sqrt{N} M_N / \hat\sigma_N,
\]
which under the hypothesis $H$ is asymptotically $N(0,1)$-distributed.
One chooses a one-sided rejection region and rejects with asymptotic
level $\alpha$ if $T_N > q_{1-\alpha}$.
If under the alternative $\hat\sigma_N$ remains bounded, we obtain
$T_N \to\infty$ in probability, so that the test is consistent.

Estimation of the long-run variance $\sigma^2$ is a delicate task.
There is a large literature starting with \citet{MR890864}, who
already propose weights in (\ref{eq:asymptest}) which guarantee
nonnegativity as well as consistency.
In our situation, one could truncate the series at the fixed prediction
window $h$ and use weights one. While this works under the hypothesis,
in our simulations a higher value of $2 h$ for the truncation with
constant weights of value one gave better power properties. Further,
since under the alternative the observations do not have mean zero, we
computed actual covariances including centering (not just second moments).

2. There is a huge econometric literature on comparing the predictive
accuracy of competing forecasts, starting with the classic paper by
\citet{Diebold1995CPA}.
For a sequence of forecasts, $\hat y_1, \ldots, \hat y_N$, and
corresponding observations $y_1, \ldots, y_N$, typically the forecast
errors $e_n = y_n - \hat y_n$ are formed, and these are inserted into a
certain loss function $l(e)$. For a competing sequence of forecasts,
$\hat z_1, \ldots, \hat z_N$, the same process is applied, leading to
$\tilde e_n = y_n - \hat z_n$. The \citeauthor{Diebold1995CPA} (DM)
test statistic is now based on analyzing the asymptotic distribution of
\[
\tilde M_N = \frac{1}{N} \sum_{n=1}^N
\bigl( l(e_n) - l(\tilde e_n) \bigr).
\]
Under stationarity assumptions on the sequences of errors $(e_n)$ and
$(\tilde e_n)$, the asymptotic distribution of $\tilde M_N$ may be
analyzed, and a $t$-statistic with a two-sided rejection region may be formed.

We note that if the forecasts $\hat y_n$ correspond to a certain
functional $T$ and a sequence of information sets, and if the scoring
function $S$ is a function in the difference $e_n$, then our test is
simply the DM test, and we analyze its behavior for asymptotically
ideal forecasts based on two distinct, ordered information sets.

As a conclusion, in our situation the DM test, performed as a one-sided
test, is a consistent test for testing the effect of increasing the
information set on the forecast of the functional. Of course, the
assumption of asymptotically ideal forecasts is a strong one. However,
if it does not hold, the test remains valid as long as the observations
and both sequences of forecasts remain stationary and the CLT still
applies [see \citet{Durrett2005PTAE}, page 416, Theorem (7.6), for
sufficient conditions], in the sense that it keeps its asymptotic level
(of course, the test is then no longer consistent).

The point of view of relating the loss function precisely to the
functional to be predicted is usually not pursued in the econometric
literature [but see \citet{MR2848512}], which is often not particularly
precise on what (meaning which functional of the predictive
distribution) is actually forecast.
Using the ``wrong'' loss function for a specific functional may result
in strongly biased results; see the example in Section~1.2 in \citet
{MR2847988}. Further, there are scoring functions which are not
functions in the linear forecast errors $e = y - \hat y$; cf. \citet
{MR2847988}.

\citet{Diebold2012CPAT} revisits the DM test and, in particular,
points out distinctions between comparing forecasting models (forecasts
arising from specific econometric models), forecasting methods [from
models but taking into account the effect of parameter estimation; see,
e.g., \citet{Giacomini2006TOCP}] or mere forecasts like in the DM test,
no matter how these were generated. Our approach is well in line with
\citeauthor{Diebold2012CPAT}, as we simply compare forecasts. If these
are (at least asymptotically) ideal, the effect of increasing the
information set on the functional may be tested consistently.

3. We conclude this subsection by remarking that the above test may be
extended to the case of proper scoring rules, in order to evaluate the
effect of increasing the information set on the complete predictive
distribution.


\section{Backtesting value at risk estimates}\label{sec:var}
The most widely used risk measure in quantitative finance is the value
at risk (VaR); see, for example, \citet{Jorion2006VART}, \citet
{Christoffersen2009VARM} or \citet{MR2175089}. Formally, this is a
(high, say, $0.99$ or $0.999$) quantile of the loss distribution.

For issuing VaR forecasts, different variations exist. Unconditional
methods base the VaR on the unconditional distribution of the risk
factors, thus using a trivial information set, while conditional
methods refer to a conditional distribution typically given the
historical data. Here, the information set may vary as well; in a
portfolio point of view it only includes the portfolio returns, while a
modeling of the individual risk factors involves a larger information
set. See also Section~\ref{sec:simappl} for further details.

Following \citet{Berkowitz2011EVAR}, typical areas of application of
VaR estimates include the following:

\textit{\textup{A.} Risk controls for trading desks.}   The distinct trading
desks (equities, currencies, derivatives, fixed-income) have limits for
the VaR, typically one-day ahead, of their trading position. These are
set by the management and monitored in real time by the back office.

\textit{\textup{B.} Portfolio choice.}  Instead of the classical Markowitz
mean--variance portfolio optimization, the VaR is used as a risk measure
when forming the optimal portfolios. Here, longer time horizons (month,
quarter) are considered, and a multivariate modeling of the risk
factors is required; see \citet{Christoffersen2009VARM}.

\textit{\textup{C.} Regulatory uses.}  Commercial banks are required to hold a
certain amount of safe assets. When based on internal methods, this
amount is determined as a function of the VaR, over a two-week horizon
and at a level of 99\%.

Different goals may be pursued for the specific VaRs reported in each
scenario. For example, in case C, the bank will be interested to report
a ``small'' (but still valid) VaR so that the required amount of
regulatory capital is reasonably small. Further, the VaR reported in
C should not vary too much over time, since the regulatory capital can
and should not be shifted abruptly.

In any of the three cases, it is of major interest to quantify and
minimize the expected amount of losses resulting from exceedences of
the VaR estimates which are being reported. To this end, our result
which relates the expected score for the VaR to the expected shortfall
is of major interest. Below we deduce from Corollary~\ref
{cor:condineq} that ideal VaR forecasts are improved in terms of the
expected shortfall arising from their exceedences by increasing the
information set.

\subsection{Exceedance indicators}\label{sec:exceedind}

Evaluating the VaR forecasts is called backtesting. In unconditional
backtesting, one checks whether the relative frequency of exceedances
of the VaR estimates corresponds to the level of the VaR; see \citet
{MR2175089}. While both unconditional and conditional methods (if
correctly specified) keep the level, the empirical level of exceedances
alone does not imply that the sequence of forecasts issued is actually
related to a quantile. Indeed, suppose that $\alpha=0.99$, then simply
issue systematically $99$ extremely high values followed by a single
extremely low value (resulting in nonstationary forecasts). This way, a
very quick convergence of the empirical exceedances to the nominal
level will be observed, but the forecasts do not make sense.

Conditional methods are often accompanied by independence checks, the
basis of which is the following well-known proposition.
For a strictly increasing, continuous distribution function $F$ let
$q_\alpha(F)$ denote the $\alpha$ quantile.
%
\begin{prop}\label{prop:independ}
Suppose that for each $\omega\in\Omega$, the conditional
distribution function $F_{Y|\F}(\omega, \cdot)$ is continuous and
strictly increasing. Let $Z$ be an $\F$-measurable random variable and
let $I = 1_{Y > Z}$ be the exceedance indicator. Then the following
assertions 1 and 2 are equivalent:

\begin{longlist}[1.]
\item[1.] $P(I=1) = 1-\alpha$, and $I$ and $\F$ are independent.

\item[2.] $Z(\omega) = q_\alpha (F_{Y|\F}(\omega, \cdot) )$ for
$P$-a.e. $\omega\in\Omega$.
\end{longlist}
\end{prop}
The proposition implies the following so-called i.i.d. and hence the
joint hypothesis [see \citet{Christoffersen1998EIF}].

\begin{cor}\label{cor:indexceed}
Suppose that $(Y_n)$ is a sequence of random variables and that $(\F
_n)$ is any filtration to which $(Y_n)$ is adapted (i.e., $Y_n$ is $\F
_n$-measurable). Suppose further that all conditional distribution
functions $F_{Y_n| \F_{n-1}}$ are continuous and strictly increasing.
Then for the one-step prediction $\hat Y_n = q_\alpha (F_{Y_n| \F
_{n-1}}(\omega, \cdot) )$, the sequence of exceedance indicators
$I_n = 1_{Y_n > \hat Y_n}$ is independent and Bernoulli distributed
with success probability $1-\alpha$.
\end{cor}

Some remarks are in order.

1. The corollary is useful for checking whether for a given sequence of
information sets, a certain forecasting method which will be based on
specification and testing works adequately. Several tests have been
proposed, taking into account effects of model misspecification and
estimation schemes; cf. \citet{Escanciano2011RBTF}.

2. However, as remarked, for example, in \citet{Escanciano2011RBTF},
mere independence of the exceedance indicators does not appropriately
take into account the role of the sequence of information sets $(\F
_n)$, since all that is needed is that $(Y_n)$ is adapted to $(\F_n)$.

3. When increasing the information sets, for example, by multivariate
modeling of risk factors, one cannot expect that the average of the
exceedance indicators will be systematically closer to the level
$1-\alpha$, which is, however, sometimes taken as a criterion [see
\citet{MR2175089}, pages 55--59]. Indeed, the speed of convergence in
$\frac{1}{N} \sum_{n=1}^N I_n \to1-\alpha$ for independent $(I_n)$
is governed by the central limit theorem
\[
\sqrt{N} \Biggl( \frac{1}{N} \sum_{n=1}^N
I_n - (1-\alpha) \Biggr) \stackrel{d} {\to} N \bigl(0, \alpha(1-
\alpha) \bigr).
\]
In order to decrease the asymptotic variance $\alpha(1-\alpha)$,
negatively correlated exceedance indicators are required, and in order
to attain a faster rate than $\sqrt{N}$, nonstationary forecasts need
to be issued as in the stylized example above.

4. The situation is even worse for $h$-step forecasts, which are
therefore comparatively rarely investigated in academic studies. Here,
$\hat Y_n^{(h)} = q_\alpha (F_{Y_n| \F_{n-h}}(\omega, \cdot
) )$, and exceedance indicators $I_n = 1_{Y_n > \hat Y_n^{(h)}}$
are only independent for lags $\geq h$.

5. In principle, the VaR based on the specific information set $\F
_{n-h}$ can be identified from the exceedance indicator by checking
full independence against the information set $\F_{n-h}$; see
Proposition~\ref{prop:independ}. Some tests take into account the
required independence of exceedance indicators to additional lagged
variables; see \citet{Berkowitz2011EVAR}. However, the question arises
what the particular additional gain is from this extended independence property.

\subsection{Quantile loss and the expected shortfall}\label{sec:quantileloss}

In what sense are ideal VaR forecasts then improved by increasing the
information set? A suitable answer seems to be provided by the theory
of the previous section, using scoring functions.

Indeed, the $\alpha$ quantile is elicitable, and the strictly
consistent scoring functions satisfying (\ref{eq:propscore}) are given by
%
\begin{equation}
\label{eq:scorequant} S(x,y) = (1_{x \geq y} - \alpha ) \bigl(g(x) - g(y) \bigr),
\end{equation}
where $g$ is strictly increasing (and all relevant expected values are
assumed to exist); see \citet{MR2847988}. Note that we can drop the
term $\alpha g(y)$ from (\ref{eq:scorequant}) and retain a strictly
consistent scoring function [though no longer nonnegative, and not
necessarily satisfying (\ref{eq:propscore})]. An attractive special
case is the choice $g(x) = x/\alpha$. After substracting $y$, we
arrive at the (no longer nonnegative) strictly consistent scoring function
\[
S^*(x,y) = \frac{1}{\alpha} 1_{x \geq y} (x-y ) - x = x \bigl(
\alpha^{-1} 1_{x \geq y} -1 \bigr) - y \alpha^{-1}
1_{x
\geq y}.
\]
Now we relate the score under $S^*$ to the expected shortfall.
%
\begin{prop}\label{prop:meansscorequant}
Suppose that $Y$ is integrable and that for each $\omega\in\Omega$
the conditional distribution function $F_{Y|\F}(\omega, \cdot)$ is
continuous and strictly increasing. For the conditional quantile $\hat
Y_\F( \omega) = q_\alpha (F_{Y| \F}(\omega, \cdot) )$ we get
%
\begin{equation}
\label{eq:meanscorequant} \qquad E \bigl(S^* (\hat Y_\F, Y )|\F \bigr) (\omega) = -
\frac
{1}{\alpha} \int_{-\infty}^{\hat Y_\F(\omega)} y F_{Y| \F
}(
\omega,dy)\qquad \mbox{for }P\mbox{-a.e. } \omega\in\Omega.
\end{equation}
Moreover, if $\F\subset\cG\subset\Aa$ and $\hat Y_\cG( \omega) =
q_\alpha (F_{Y| \cG}(\omega, \cdot) )$, then
\begin{eqnarray}
\label{eq:increaseinfoquant} 
&&- \frac{1}{\alpha} \int_{-\infty}^{\hat Y_\cG(\omega)}
y F_{Y|
\cG}(\omega,dy) \nonumber\\
&&\qquad \leq
- \frac{1}{\alpha} \int
_{-\infty}^{\hat Y_\F(\omega)} y F_{Y| \F}(\omega,dy)\qquad \mbox{for
}P\mbox{-a.e. } \omega\in \Omega,
\nonumber\\
&&E \biggl(- \frac{1}{\alpha} \int_{-\infty}^{\hat Y_\cG(\cdot)}
y F_{Y| \cG}(\cdot,dy) \Big| \F \biggr) (\omega)
\nonumber
\\[-8pt]
\\[-8pt]
\nonumber
&&\qquad\leq -
\frac{1}{\alpha} \int_{-\infty}^{\hat Y_\F(\omega)} y F_{Y|
\F}(
\omega,dy)\qquad \mbox{for }P\mbox{-a.e. } \omega\in\Omega,
\\
&&\int_\Omega- \frac{1}{\alpha} \int
_{-\infty}^{\hat Y_\cG(\omega
)} y F_{Y| \cG}(\omega,dy) \,dP(\omega)
\nonumber\\
&&\qquad
 \leq\int_\Omega- \frac{1}{\alpha} \int
_{-\infty}^{\hat Y_\F(\omega
)} y F_{Y| \F}(\omega,dy) \,dP(\omega),\nonumber
\end{eqnarray}
with equality in one of the inequalities in (\ref
{eq:increaseinfoquant}) if and only if $\hat Y_\cG= \hat Y_\F$ a.s.
\end{prop}
For an interpretation, suppose that $Y$ corresponds to the profit and
loss distribution (e.g., is a log-return), so that $\alpha$ is indeed
a small value such as $\alpha= 0.01$ or $0.001$.
Then
\[
- \frac{1}{\alpha} \int_{-\infty}^{\hat Y_\F(\omega)} y
F_{Y|
\F}(\omega,dy)
\]
is the lower-tail expected shortfall of the conditional distribution
and, thus, $E (S^*(\hat Y_\F( \omega), Y)  )$ as in (\ref
{eq:increaseinfoquant}) is the mean lower-tail expected shortfall when
using the information set $\F$. \citet{Rockafellar2000OOCV} give a
result similar to~(\ref{eq:meanscorequant}); see their Theorem~1.
%

%

%
\section{Simulations and applications}\label{sec:simappl}
In this section we investigate the proposed methods in the context of
value at risk estimation both in simulated examples as well as for
log-returns of several stocks and stock indices. We let $T\dvtx \Theta\to
\R$ be the $\alpha$ quantile, and let $S(x,y) = x   ( \alpha
^{-1} 1_{x \geq y} -1 ) - y  \alpha^{-1} 1_{x \geq y}$; see
Section~\ref{sec:var}. While the quantile loss function has been used
in some numerical studies [cf. \citet{MR2226780}], the particular
effect of the information set does not seem to have been investigated
so far.
%
\subsection{Unconditional versus conditional risk management}
We consider the situation of conditional versus unconditional risk
management; see \citet{MR2175089}. Let $(R_t)_{t \in\Z}$ be a
stationary time series corresponding to daily log-returns of a stock or
stock index, and let
\[
\F_t = \{ \varnothing, \Omega\},\qquad \cG_t= \sigma
\{R_s\dvtx s \leq t\}.
\]
Thus, forecasts based on the trivial $\F_t$ concern the unconditional
distribution of returns, while forecasts based on $\cG_t$ concern the
conditional distribution given daily log-returns. Fix some prediction
horizon $h \geq1$, and set
\[
Y_{t+h} = Y_{t+h}^{(h)} = R_{t+1} + \cdots+
R_{t+h},
\]
the $h$-step log-return. Our aim is $h$-step forecasting of the
quantile of $Y_t$, that is,
\[
\hat Y_{t+h,\F}^{(h)} = T ( F_{Y_{t+h}|\F_{t} } ) \quad\mbox{and}\quad \hat
Y_{t+h,\cG}^{(h)} = T ( F_{Y_{t+h}|\cG
_{t} } ).
\]
Since $\F_t$ is trivial, the $\hat Y_{t+h,\F}^{(h)}$ are constant
a.s. and equal to the unconditional quantile of the $Y_t$, while $\hat
Y_{t+h,\cG}^{(h)}$ is the conditional quantile of the $h$-step return
given the history of one-step returns up to time $t$.
For the conditional method, the exceedance indicators $1_{\hat
Y_{t+h,\cG}^{(h)} > Y_{t+h}}$ are independent for lags $\geq h$, while
there is no such general independence for the unconditional method.
However, note that for larger values of $h$, it is quite hard to
distinguish both methods based on (non) independence.

\textit{Simulation}.
As a data-generating process, we use a $\operatorname{GARCH}(1,1)$-model
\[
R_{t} = \sigma_{t} \varepsilon_t,\qquad
\sigma_t^2 = \kappa+ \phi R_{t-1}^2
+ \beta\sigma_{t-1}^2,
\]
where the $(\varepsilon_t)$ are i.i.d. $N(0,1)$-distributed, and the
distinct parameter values $(\kappa, \phi, \beta)$ are chosen
according to the scenarios in Table~\ref{Tab-S-U-config}.
As prediction horizons we consider $h=1, 2$: one and two days, $h=10$:
two weeks, $h=66$: one quarter of the year. Given the parameters of the
GARCH model (either true values or estimates) as well as estimates of
the one-step volatilities $\sigma_{t}^2$, as conditional forecasts we
use in case $h=1$ the exact forecast distribution $N(0, \sigma
_{t+1}^2)$, while for $h >1$ we approximate the quantile by the
empirical quantile of a Monte Carlo sample of size $M=1000$ for each
$t$. As an unconditional forecast we use an $\alpha$ quantile of an
appropriate series of $h$-step returns.

(a) First, we briefly investigate the true expected mean scores for
unconditional and conditional risk management using (approximate) ideal
forecasts, which by (\ref{eq:increaseinfoquant}) correspond to average
expected shortfalls. To this end, we use a single huge sample of size
$N=100{,}000$ (resp.,  $N=300{,}000$ for $h=1$). For the conditional
forecasts $\hat Y_{t+h,\cG}^{(h)}$, we use the true parameters of the
GARCH model, while for the unconditional case, we set $\hat Y_{t+h,\F
}^{(h)}$ constant as the empirical quantile of a distinct simulated
series of $(Y_t)$ of length $300{,}000$. Finally, we approximate the
mean score by the sample averages $\widehat m_{N, \F}$ and $\widehat
m_{N, \cG}$ as in (\ref{eq:converg}).

The results for configuration 1 can be found in Table~\ref{Tab-S-U-g};
for the other configurations these are similar. As stated in
\citeauthor{Acerbi2002OTCO} [(\citeyear{Acerbi2002OTCO}),
Proposition~3.4], we see that for fixed $h$ and
increasing values of $\alpha$, the values of $\widehat m_{N, \F}$ and
$\widehat m_{N, \cG}$ decrease.
Moreover, for fixed $\alpha$ and increasing values of $h$, $\widehat
m_{N, \F}$ and $\widehat m_{N, \cG}$ increase. The relative
difference, $M_N/\hat\mu_{N, \F}$, which indicates the reduction in
mean expected shortfall when passing from the unconditional to the
conditional method, is highest for small $\alpha$ for fixed $h$, with
values as large as $31\%$.

The estimate $\hat\sigma^2$ for $\sigma^2 = E(Z_1^2 + 2 \sum_{k=2}^\infty Z_1 Z_k)$, where the $Z_k$ are as in (\ref
{eq:asymptest}), is obtained by truncation at $2h$ with constant weight
one, and where the observations are centered before computing
covariances. This choice gave reasonable power properties in our simulations.

The last column contains the values $T_N$ of the $t$-statistic together
with the $p$-value based on the asymptotic approximation.
For the values $h=1,2$ and $10$, the difference is significantly $>$0
for all $\alpha$, while for $h= 66$ it is not significant.

%
\begin{table}
\tablewidth=250pt
\caption{Parameter configurations for the $\operatorname{GARCH}(1,1)$ model in the
comparison of conditional versus unconditional VaR estimation}
\label{Tab-S-U-config}
\begin{tabular*}{250pt}{@{\extracolsep{\fill}}lcd{1.3}d{1.3}@{}}
\hline
\textbf{Config.} & \multicolumn{1}{c}{$\bolds{\kappa}$} & \multicolumn{1}{c}{$\bolds{\phi}$} &
\multicolumn{1}{c@{}}{$\bolds{\beta}$} \\
\hline
1 & 0.01 & 0.088 & 0.902 \\
2 & 0.02 & 0.2 & 0.78 \\
3 & 0.05 & 0.3 & 0.65 \\
\hline
\end{tabular*}
\end{table}


%
\begin{table}
\caption{Mean scores for conditional and unconditional VaR estimation
for parameter configuration 1, see Table \protect\ref{Tab-S-U-config}}
\label{Tab-S-U-g}
\begin{tabular*}{\textwidth}{@{\extracolsep{\fill}}lcd{2.3}d{2.3}ccd{3.1}d{2.1}c@{}}
\hline
& &
\multicolumn{2}{c}{\textbf{Mean scores}} & \multicolumn
{1}{c}{\textbf{Diff.} $\bolds{(=M_N)}$} &
\multicolumn{1}{c}{\textbf{Rel. diff.}}
& & \\[-6pt]
& &
\multicolumn{2}{c}{\hrulefill} & \multicolumn
{1}{c}{\hrulefill} & \multicolumn{1}{c}{\hrulefill}
& & \\
\multicolumn{1}{@{}l}{$\bolds{h}$} & \multicolumn{1}{c}{$\bolds{\alpha}$} &
\multicolumn{1}{c}{$\bolds{\hat m_{N, \F}}$} & \multicolumn{1}{c}{$\bolds{\hat
m_{N, \cG}}$} & \multicolumn{1}{c}{$\bolds{\hat m_{N, \F} - \hat m_{N, \cG
}}$} & \multicolumn{1}{c}{$\bolds{M_N/\hat m_{N, \F}}$} & \multicolumn
{1}{c}{$\bolds{\hat\sigma}$} & \multicolumn{1}{c}{$\bolds{T_N}$} & \multicolumn
{1}{c@{}}{$\bolds{\operatorname{Pr}(>\!T_N)} $} \\
\hline
\phantom{0}1 & 0.01 & 3.627 & 2.511 &
1.116 & 0.31 & 15.7 & 38.9 & $<$0.001 \\
\phantom{0}1 & 0.05 & 2.225 & 1.895 & 0.330 & 0.15 & 3.4 & 52.4 & $<$0.001 \\
\phantom{0}1 & 0.20 & 1.354 & 1.303 & 0.051 & 0.04 & 0.7 & 40.5 & $<$0.001 \\[3pt]
\phantom{0}2 & 0.01 & 4.547 & 3.573 & 0.974 & 0.21 & 18.9 & 8.1 & $<$0.001 \\
\phantom{0}2 & 0.05 & 3.652 & 3.035 & 0.617 & 0.17 & 8.0 & 12.3 & $<$0.001 \\
\phantom{0}2 & 0.20 & 1.882 & 1.828 & 0.055 & 0.03 & 1.2 & 7.5 & $<$0.001 \\[3pt]
10 & 0.01 & 12.852 & 9.579 & 3.272 & 0.25 & 115.8 & 4.5 & $<$0.001 \\
10 & 0.05 & 6.749 & 5.988 & 0.761 & 0.11 & 19.5 & 6.2 & $<$0.001 \\
10 & 0.20 & 3.991 & 3.890 & 0.101 & 0.03 & 4.6 & 3.5 & $<$0.001 \\[3pt]
66 & 0.01 & 25.331 & 24.746 & 0.585 & 0.02 & 320.9 & 0.3 & \phantom{$>$}0.387
\\
66 & 0.05 & 17.726 & 17.398 & 0.328 & 0.02 & 76.4 & 0.7 & \phantom{$>$}0.249 \\
66 & 0.20 & 11.552 & 11.407 & 0.145 & 0.01 & 28.2 & 0.8 & \phantom{$>$}0.209 \\
\hline
\end{tabular*}
\end{table}

%

(b) Next, we investigate the power of the resulting DM test for
realistic sample sizes when taking into account estimation effects.
We based estimation of the parameters of the GARCH model for the
unconditional method as well as of the quantile for the unconditional
method on a rolling window of size $R_\mathrm{wind} = 500$. For the
unconditional method, we investigated two variations, first using the
empirical quantile of the last $R_\mathrm{wind}$ $h$-step returns
preceding $t$, and second using a square root of time rule resulting in
$\hat Y_{t+h}^{(h)} = \sqrt{h} \hat s_t q_\alpha+ h \hat m_t$, where
$\hat s_t$ and $\hat m_t$ are the empirical standard deviation and mean
of the last $R_\mathrm{wind}$ one-step returns preceding $t$ and
$q_\alpha$ is the $\alpha$ quantile of the standard normal. Since the
square root of time rule in most cases led to smaller scores, we only
displayed the corresponding results. Note that due to the limited
estimation horizon, the unconditional method is in fact also partially
conditional. We then compute the DM $t$-statistic $T_N$ with the estimate
for the long-run variance as described above. This is iterated $1000$ times.

Results for the three configurations of Table~\ref{Tab-S-U-config},
various sample sizes $N$ (so that the number of observations is $N +
R_\mathrm{wind}$), test levels $0.05$ and $0.1$ and $h=1,2,10$ are
displayed in Tables~\ref{Tab-S-U-k-point05} and \ref
{Tab-S-U-k-point1}. For $h=66$, the test does not have any power beyond
the level. Otherwise, the power properties are quite reasonable.

\begin{table}
\caption{Power of the test (at the $0.05$ level) for conditional and
unconditional VaR estimation ($\alpha= 0.01$); for parameter
configurations, cf. Table \protect\ref{Tab-S-U-config}}
\label{Tab-S-U-k-point05}
\begin{tabular*}{\textwidth}{@{\extracolsep{\fill}}lcccc@{}}
\hline
 & & \multicolumn{3}{c@{}}{\textbf{Config.}} \\[-6pt]
& & \multicolumn{3}{c@{}}{\hrulefill} \\
\multicolumn{1}{@{}l}{$\bolds{h}$} & $\bolds{N}$ & \multicolumn{1}{c}{$\bolds{1}$} &
\multicolumn{1}{c}{$\bolds{2}$} & \multicolumn{1}{c}{$\bolds{3}$} \\
\hline
\phantom{0}{1} & \phantom{0}250 & 0.463 & 0.565 & 0.479 \\
& \phantom{0}500 & 0.632 & 0.776 & 0.640 \\
& 1000 & 0.863 & 0.951 & 0.900 \\
& 1500 & 0.947 & 0.993 & 0.981 \\[3pt]
\phantom{0}{2} & \phantom{0}250 & 0.392 & 0.421 & 0.326 \\
& \phantom{0}500 & 0.492 & 0.576 & 0.447 \\
& 1000 & 0.723 & 0.844 & 0.744 \\
& 1500 & 0.859 & 0.957 & 0.905 \\
& 2000 & 0.920 & 0.984 & 0.970 \\
& 4000 & 0.999 & 0.999 & 0.999 \\[3pt]
{10} & \phantom{0}250 & 0.258 & 0.214 & 0.140 \\
& \phantom{0}500 & 0.196 & 0.157 & 0.087 \\
& 1000 & 0.205 & 0.173 & 0.079 \\
& 1500 & 0.277 & 0.232 & 0.119 \\
& 2000 & 0.330 & 0.306 & 0.162 \\
& 4000 & 0.634 & 0.620 & 0.412 \\
\hline
\end{tabular*}
\end{table}

%
\begin{table}
\caption{Power of the test (at the $0.1$ level) for conditional and
unconditional VaR estimation ($\alpha= 0.01$); for parameter
configurations, cf. Table \protect\ref{Tab-S-U-config}}
\label{Tab-S-U-k-point1}
\begin{tabular*}{\textwidth}{@{\extracolsep{\fill}}lcccc@{}}
\hline
 & & \multicolumn{3}{c@{}}{\textbf{Config.}} \\[-6pt]
& & \multicolumn{3}{c@{}}{\hrulefill} \\
\multicolumn{1}{@{}l}{$\bolds{h}$} & $\bolds{N}$ & \multicolumn{1}{c}{$\bolds{1}$} &
\multicolumn{1}{c}{$\bolds{2}$} & \multicolumn{1}{c}{$\bolds{3}$} \\
\hline
\phantom{0}{1} & \phantom{0}250 & 0.578 & 0.693 & 0.629 \\
& \phantom{0}500 & 0.757 & 0.870 & 0.792 \\
& 1000 & 0.921 & 0.982 & 0.956 \\
& 1500 & 0.978 & 0.998 & 0.990 \\[3pt]
\phantom{0}{2} & \phantom{0}250 & 0.523 & 0.612 & 0.510 \\
& \phantom{0}500 & 0.666 & 0.766 & 0.688 \\
& 1000 & 0.858 & 0.937 & 0.912 \\
& 1500 & 0.934 & 0.987 & 0.981 \\
& 2000 & 0.974 & 0.995 & 0.991 \\
& 4000 & 1.000 & 1.000 & 1.000 \\[3pt]
{10} & \phantom{0}250 & 0.350 & 0.288 & 0.206 \\
& \phantom{0}500 & 0.335 & 0.295 & 0.189 \\
& 1000 & 0.416 & 0.372 & 0.233 \\
& 1500 & 0.509 & 0.495 & 0.321 \\
& 2000 & 0.590 & 0.597 & 0.410 \\
& 4000 & 0.819 & 0.842 & 0.689 \\
\hline
\end{tabular*}
\end{table}

\textit{Application}.
Finally, we investigate unconditional versus conditional risk
management when applied to log-returns of several stocks and stock-indices.
We use publicly available share prices of German stocks (on a daily
basis) from Yahoo Finance
(\url{http://finance.yahoo.com}). The data set runs from 1st January $2001$
to 31st July $2013$. In the direct comparison of two shares we restrict for
simplicity to the subset of available data points (for each share) by
taking intersections. In any case, the subset of share prices in our
analysis was larger than $2727$ (each including the beginning of the
year $2003$). Let $S_t$ denote the price, $R_t = \log S_t - \log
S_{t-1}$ the log-return, so that
\[
Y_{t+h}^{(h)} = \log S_{t+h} - \log S_t
\]
is the $h$-step log-return. We proceed as in the simulations part
(b) above, using a rolling window of size $500$ as well as the
square-root-of-time rule for the unconditional method.
The results for various stocks can be found in Table~\ref{Tab-A-U}.
The mean score is significantly reduced for $h=1$ and $h=2$ when
passing from the unconditional to the conditional methods, where the
maximal value for the relative difference is $0.3$. For higher lags,
the reduction is nonsignificant.

%
\begin{table}
\tabcolsep=0pt
\caption{Mean scores for conditional and unconditional VaR estimation
($\alpha= 0.01$) for the log-returns of several stocks (date values
starting from at least 2001-01-02 resulting in a value of $N\geq3211$
in each row)}
\label{Tab-A-U}
\begin{tabular*}{\textwidth}{@{\extracolsep{\fill}}ld{2.0}d{2.2}d{2.2}ccd{3.1}cc@{}}
\hline
 & & \multicolumn{2}{c}{\textbf{Mean
scores}} & \multicolumn{1}{c}{\textbf{Diff.} $\bolds{(=M_N)}$} & \multicolumn
{1}{c}{\textbf{Rel. diff.}} & & \\[-6pt]
& & \multicolumn{2}{c}{\hrulefill} & \multicolumn{1}{c}{\hrulefill} & \multicolumn
{1}{c}{\hrulefill} & & \\
\multicolumn{1}{@{}l}{\textbf{Share name}}
& \multicolumn{1}{c}{$\bolds{h}$} & \multicolumn{1}{c}{$\bolds{\hat m_{N, \F}}$} &
\multicolumn{1}{c}{$\bolds{\hat m_{N, \cG}}$} & \multicolumn{1}{c}{$\bolds{ \hat
m_{N, \F} - \hat m_{N, \cG}}$} & \multicolumn{1}{c}{$\bolds{M_N/\hat m_{N,
\F}}$} & \multicolumn{1}{c}{$\bolds{\hat\sigma}$} & \multicolumn
{1}{c}{$\bolds{T_N}$} & \multicolumn{1}{c}{$\bolds{\operatorname{Pr}(>\!T_N)}$} \\
\hline
{DAX} & 1 & 5.87 & 4.24 & 1.63 & 0.28 &
15.4 & 5.48 & 0.000 \\
& 2 & 8.43 & 6.46 & 1.97 & 0.23 & 29.3 & 3.50 & 0.000 \\
& 10 & 22.07 & 18.26 & 3.80 & 0.17 & 67.8 & 2.91 & 0.002 \\[3pt]
{Daimler} & 1 & 8.62 & 6.92 & 1.70 & 0.20
& 18.6 & 4.81 & 0.000 \\
& 2 & 12.14 & 9.75 & 2.39 & 0.20 & 36.6 & 3.42 & 0.000 \\
& 10 & 34.44 & 29.44 & 5.00 & 0.15 & 193.2 & 1.35 & 0.088 \\[3pt]
{Deutsche Bank} & 1 & 10.08 & 7.19 & 2.89
& 0.29 & 40.9 & 3.71 & 0.000 \\
& 2 & 15.89 & 10.90 & 4.99 & 0.31 & 85.2 & 3.08 & 0.001 \\
& 10 & 38.56 & 29.39 & 9.17 & 0.24 & 450.7 & 1.06 & 0.144 \\[3pt]
{Munich RE} & 1 & 7.49 & 6.06 & 1.42 &
0.19 & 17.9 & 4.14 & 0.000 \\
& 2 & 10.92 & 8.75 & 2.17 & 0.20 & 35.0 & 3.22 & 0.001 \\
& 10 & 21.17 & 19.92 & 1.25 & 0.06 & 124.4 & 0.52 & 0.300 \\[3pt]
{Siemens} & 1 & 8.89 & 6.85 & 2.04 & 0.23
& 28.5 & 3.75 & 0.000 \\
& 2 & 12.07 & 9.32 & 2.75 & 0.23 & 39.0 & 3.70 & 0.000 \\
& 10 & 31.33 & 26.09 & 5.24 & 0.17 & 95.7 & 2.87 & 0.002 \\
\hline
\end{tabular*}
\end{table}

\textit{Conclusions}.
For $h=1$ and $h=2$, the improved performance of the conditional method
compared to the unconditional method is apparent, both in the
simulations and also in the stock returns. On the other hand, for
$h=66$ (the quarter) there is no significant improvement for the stock
returns, and the potential improvement as indicated by the simulations
is also small. For $h=10$ (two weeks), simulations indicate quite a
potential for improvement, but the effect in the actual stock returns,
if present, is often not yet significant.

\subsection{Univariate versus multivariate modeling for risk management}

Now we consider a univariate modeling on the basis of portfolio returns
versus a multivariate modeling of the individual risk factors.
For simplicity we only investigate two underlying risk factors.

Let
$(\bR_t)_{t \in\Z}$, $\bR_t = (R_{t,1}, R_{t,2})^T$ be a stationary
bivariate time series corresponding to daily returns of the individual
stocks of a portfolio. For a fixed weight vector $\bw= (w_1, w_2)^T$,
with $0\leq w_i \leq1$, $w_1 + w_2=1$, we let $Y_t = \bw^T \bR_t$,
which we interpret as the return of a portfolio consisting of the two
individual stocks. Note that on the basis of the prices of the
portfolio, this corresponds to a reweighting in each step; see the
application below. As information sets, consider
\[
\F_t = \sigma\{Y_s\dvtx s \leq t\},\qquad
\cG_t= \sigma\{\bR _s\dvtx s \leq t\},
\]
the history of portfolio returns $\F_t$ and of individual risk factors
$\cG_t$. Our aim is one-step forecasting of the quantile of $Y_t$,
that is,
\[
\hat Y_{t+1,\F}^{(1)} = \hat Y_{t+1,\F}= T (
F_{Y_{t+1}|\F_{t}
} ) \quad\mbox{and}\quad \hat Y_{t+1,\cG}^{(1)}= \hat
Y_{t+1,\cG} = T ( F_{Y_{t+1}|\cG_{t} } ).
\]
Thus, $\hat Y_{t+1,\F}$ is the forecast based on the history of
portfolio returns, while $\hat Y_{t+1,\cG}$ is the forecast based on
the history of individual risk factors. Note that in both cases, for
ideal forecasts the series of exceedance indicators $(I_{t,\F})$ and
$(I_{t, \cG})$, where $I_{t,\F} = 1_{ \hat Y_{t, \F} > Y_t}$ and
$I_{t,\cG} = 1_{ \hat Y_{t, \cG} > Y_t}$, are both
Bernoulli-sequences with success probabilities $\alpha$.\\
\textit{Simulation}
We simulate the series $(\bR_t)$ from a bivariate DCC-GARCH-model of
\citet{MR1939905}, where
%
\begin{eqnarray}
\bR_t & =& H_t^{1/2}
\bepsilon_t \qquad\mbox{with } \bepsilon_t \mbox{ i.i.d.}
\sim N(\bo, I_2),
\nonumber\\
H_t & = &D_t C_t
D_t,\qquad  D_t = \operatorname{diag} (\sigma_{t,1},
\sigma_{t,2}),
\nonumber\\
\sigma_{t,i}^2 & =&
\kappa_i + \phi_i^2 R_{t-1,i}^2
+ \beta_i \sigma_{t,i-1}^2,
\nonumber
\\[-8pt]
\\[-8pt]
\nonumber
\qquad C_t &=& \operatorname{diag}
\bigl(q_{t;1,1}^{-1/2},q_{t;2,2}^{-1/2} \bigr)
Q_t \operatorname{diag} \bigl(q_{t;1,1}^{-1/2},q_{t;2,2}^{-1/2}
\bigr),\qquad  Q_t = (q_{t;j,k} )_{j,k=1,2},
\\
\nonumber
Q_t & =& (1 - \gamma- \eta) \bar Q + \gamma
\bu_{t-1} \bu_{t-1}^T + \eta Q_{t-1},
\\
\nonumber
\bu_t & = &(R_{t,1}/
\sigma_{t,1}, R_{t,2}/\sigma_{t,2} )^T,\qquad
\bar Q = \operatorname{cov} (\bu_t), %
\end{eqnarray}
and the parameters are chosen according to the scenarios listed in
Table~\ref{Tab-S-M-DCC-config}, $\bw=(1/2, 1/2)^T$ and $\alpha=0.01$.

(a) Again, we first approximate the true mean score of the
(approximate) ideal forecasts by sample averages $\widehat m_{N, \cG}$
and $\widehat m_{N, \F}$ based on a single huge sample of size
$N=500{,}000$. In the multivariate case for $\hat Y_{t+1,\cG}$ and
$\widehat m_{N, \cG}$, we use the true parameters of the
DCC-GARCH-model and the exact forecast distribution $N(0, \bw^T H_t
\bw)$. For $\hat Y_{t+1,\F}$ and $\hat\mu_{N, \F}$, we first
determine an appropriate model for the series of $(Y_t)$ within the
class of $\operatorname{GARCH}(p,q)$-models, and then use one-step forecasts within
this univariate GARCH-model.
Even though the class of multivariate GARCH models is not closed under
aggregation, it turns out that a simple $\operatorname{GARCH}(1,1)$-model with normal
innovations works surprisingly well.
The results can be found in Table~\ref{Tab-S-M-DCC-g}. While in all
scenarios the difference between the average scores is significant due
to the high sample sizes, the relative reduction in mean score is small
with maximal values of $0.06$.

Simulations for a class of regime-switching models which are closed
under aggregation led to similar results.

%
\begin{table}
\caption{Configurations for the simulation of the DCC-GARCH-model
($N=500{,}000$, $\alpha=0.01$, $w_1=0.5$, $w_2=0.5$)}
\label{Tab-S-M-DCC-config}
\begin{tabular*}{\textwidth}{@{\extracolsep{\fill}}lccccccccc@{}}
\hline
\multicolumn{1}{@{}l}{\textbf{Config.}} & \multicolumn{1}{c}{$\bolds{\kappa_1}$} &
\multicolumn{1}{c}{$\bolds{\kappa_2}$} & \multicolumn{1}{c}{$\bolds{\phi_1}$} &
\multicolumn{1}{c}{$\bolds{\phi_2}$} & \multicolumn{1}{c}{$\bolds{\beta_1}$} &
\multicolumn{1}{c}{$\bolds{\beta_2}$} & \multicolumn{1}{c}{$\bolds{\bar q_{21}}$} &
\multicolumn{1}{c}{$\bolds{\gamma}$} & \multicolumn{1}{c@{}}{$\bolds{\eta}$} \\
\hline
1 & 0.0030 & 0.0010 & 0.400 & 0.050 & 0.590 & 0.930 & 0.10 & 0.01 &
0.98 \\
2 & 0.0025 & 0.0015 & 0.390 & 0.060 & 0.600 & 0.920 & 0.30 & 0.02 &
0.97 \\
3 & 0.0100 & 0.0070 & 0.200 & 0.180 & 0.790 & 0.800 & 0.30 & 0.08 &
0.91 \\
4 & 0.0200 & 0.0010 & 0.100 & 0.300 & 0.890 & 0.680 & 0.35 & 0.10 &
0.89 \\
5 & 0.0030 & 0.0010 & 0.400 & 0.005 & 0.590 & 0.975 & 0.60 & 0.01 &
0.98 \\
6 & 0.0090 & 0.0080 & 0.200 & 0.010 & 0.790 & 0.970 & 0.75 & 0.05 &
0.94 \\
7 & 0.0028 & 0.0031 & 0.300 & 0.500 & 0.690 & 0.480 & 0.88 & 0.01 &
0.98 \\
\hline
\end{tabular*}

\end{table}

%
\begin{table}[b]
\caption{Mean scores for univariate and multivariate VaR estimation
($N=500{,}000$, $\alpha=0{.}01$); for parameter configurations 1 to 7,
cf. Table~\protect\ref{Tab-S-M-DCC-config}}
\label{Tab-S-M-DCC-g}
\begin{tabular*}{\textwidth}{@{\extracolsep{\fill}}lcccccd{2.2}c@{}}
\hline
 & \multicolumn{2}{c}{\textbf{Mean scores}} &
\multicolumn{1}{c}{\textbf{Diff.} $\bolds{(=M_N)}$} &
\multicolumn{1}{c}{\textbf{Rel. diff.}} &
& \\[-6pt]
 & \multicolumn{2}{c}{\hrulefill} &
\multicolumn{1}{c}{\hrulefill} &
\multicolumn{1}{c}{\hrulefill} &
& \\
{\textbf{Config}.}& \multicolumn{1}{c}{$\bolds{ \hat{m}_{N, \F}}$} &
\multicolumn{1}{c}{$
\bolds{\hat{m}_{N, \cG}}$} & \multicolumn{1}{c}{$ \bolds{\hat{m}_{N, \F} - \hat
{m}_{N, \cG}}$} & \multicolumn{1}{c}{$\bolds{M_N/\hat{m}_{N, \F}}$} &
\multicolumn{1}{c}{$\bolds{\hat\sigma}$} & \multicolumn{1}{c}{$\bolds{T_N}$} &
\multicolumn{1}{c@{}}{$\bolds{\operatorname{Pr}(>\!T_N)}$} \\
\hline
1 & 0.527 & 0.495 &
0.031 & 0.06 & 1.1 & 20.93 & $<$0.001 \\
2 & 0.580 & 0.556 & 0.024 & 0.04 & 0.9 & 19.02 & $<$0.001 \\
3 & 1.330 & 1.322 & 0.007 & 0.01 & 0.6 & 9.18 & $<$0.001 \\
4 & 1.727 & 1.725 & 0.002 & 0.00 & 0.3 & 4.96 & $<$0.001 \\
5 & 0.595 & 0.574 & 0.021 & 0.04 & 1.2 & 12.73 & $<$0.001 \\
6 & 1.648 & 1.628 & 0.020 & 0.01 & 1.3 & 11.12 & $<$0.001 \\
7 & 0.666 & 0.662 & 0.003 & 0.00 & 0.3 & 6.33 & $<$0.001 \\
\hline
\end{tabular*}
\end{table}

%
\begin{table}
\caption{Power of the test (at the $0.05$ level) for univariate and
multivariate VaR estimation ($\alpha= 0.01$); for parameter
configurations, cf. Table \protect\ref{Tab-S-M-DCC-config}}
\label{Tab-S-M-DCC-k-point05}
\begin{tabular*}{\textwidth}{@{\extracolsep{\fill}}lccccccc@{}}
\hline
& \multicolumn{7}{c@{}}{\textbf{Config.}} \\[-6pt]
& \multicolumn{7}{c}{\hrulefill} \\
\multicolumn{1}{@{}l}{$\bolds{N}$} & \multicolumn{1}{c}{$\bolds{1}$} & \multicolumn
{1}{c}{$\bolds{2}$} & \multicolumn{1}{c}{$\bolds{3}$} & \multicolumn{1}{c}{$\bolds{4}$} &
\multicolumn{1}{c}{$\bolds{5}$} & \multicolumn{1}{c}{$\bolds{6}$} & \multicolumn
{1}{c@{}}{$\bolds{7}$} \\
\hline
\phantom{0}250 & 0.099 & 0.086 & 0.091 & 0.094 & 0.080 & 0.088 & 0.044 \\
\phantom{0}500 & 0.112 & 0.084 & 0.073 & 0.051 & 0.075 & 0.083 & 0.055 \\
1000 & 0.159 & 0.110 & 0.037 & 0.044 & 0.111 & 0.092 & 0.058 \\
1500 & 0.232 & 0.183 & 0.059 & 0.045 & 0.172 & 0.129 & 0.088 \\
2000 & 0.305 & 0.219 & 0.076 & 0.048 & 0.201 & 0.140 & 0.085 \\
4000 & 0.567 & 0.418 & 0.100 & 0.038 & 0.403 & 0.245 & 0.104 \\
6000 & 0.707 & 0.528 & 0.116 & 0.035 & 0.545 & 0.322 & 0.104 \\
\hline
\end{tabular*}
\end{table}

%
\begin{table}[b]
\caption{Power of the test (at the $0.1$ level) for univariate and
multivariate VaR estimation ($\alpha= 0.01$); for parameter
configurations, cf. Table \protect\ref{Tab-S-M-DCC-config}}
\label{Tab-S-M-DCC-k-point1}
\begin{tabular*}{\textwidth}{@{\extracolsep{\fill}}lccccccc@{}}
\hline
& \multicolumn{7}{c@{}}{\textbf{Config.}} \\[-6pt]
& \multicolumn{7}{c}{\hrulefill} \\
\multicolumn{1}{@{}l}{$\bolds{N}$} & \multicolumn{1}{c}{$\bolds{1}$} & \multicolumn
{1}{c}{$\bolds{2}$} & \multicolumn{1}{c}{$\bolds{3}$} & \multicolumn{1}{c}{$\bolds{4}$} &
\multicolumn{1}{c}{$\bolds{5}$} & \multicolumn{1}{c}{$\bolds{6}$} & \multicolumn
{1}{c@{}}{$\bolds{7}$} \\
\hline
\phantom{0}250 & 0.195 & 0.177 & 0.162 & 0.143 & 0.169 & 0.159 & 0.106 \\
\phantom{0}500 & 0.262 & 0.208 & 0.143 & 0.090 & 0.181 & 0.166 & 0.131 \\
1000 & 0.327 & 0.247 & 0.104 & 0.091 & 0.251 & 0.196 & 0.139 \\
1500 & 0.419 & 0.341 & 0.140 & 0.107 & 0.308 & 0.234 & 0.161 \\
2000 & 0.496 & 0.382 & 0.158 & 0.103 & 0.358 & 0.266 & 0.155 \\
4000 & 0.737 & 0.589 & 0.181 & 0.096 & 0.554 & 0.399 & 0.168 \\
6000 & 0.835 & 0.706 & 0.206 & 0.093 & 0.710 & 0.468 & 0.184 \\
\hline
\end{tabular*}
\end{table}

(b) Next, we investigate the power of the resulting DM test for
realistic sample sizes when taking into account estimation effects.
Again, we base estimation on a rolling window of sizes $R_\mathrm
{wind}=500$ and proceed as in part (b) above. The resulting power
estimates for test levels $0.05$ and $0.1$, which are reasonably high
at least for higher sample sizes, can be found in Tables~\ref{Tab-S-M-DCC-k-point05} and \ref{Tab-S-M-DCC-k-point1}.

\textit{Application}.
We proceed with an application to portfolios consisting of two stocks.
Let $S_{t,i}$, $i=1,2$, denote the price of stock $i$ at time $t$
(daily closure). Consider the relative returns
\[
R_{t,i} = \frac{S_{t,i} - S_{t-1,i}}{S_{t-1,i}},\qquad i=1,2.
\]
Let $\lambda_{t,i}$ denote the amount held from stock $i$ from time
$t$ to time $t+1$, and let
$ V_t = \lambda_{t,1}  S_{t,1} + \lambda_{t,2}  S_{t,2}$.
Then for the portfolio return $(Y_t)$,
\[
Y_{t+1} = \sum_{i=1}^2
R_{t+1,i} \lambda_{t,i} \frac
{S_{t,i}}{V_t},
\]
so that in order to obtain the constant weights $w_i$, $i=1,2$, on the
basis of returns, we choose
$ \lambda_{t,i} = w_i V_t/S_{t,i}$
with initial value $V_0=1$. We model the series $(\bR_t)$, $\bR_t =
(R_{t,1},R_{t,2})^T$, by a DCC-GARCH-model as specified above and the
univariate series $(Y_t)$ of portfolio returns by a simple
$\operatorname{GARCH}(1,1)$-model. In both cases, at time $t$ using a rolling window
we base the estimation on the last $R_{\mathrm{wind}}=500$ observations.

The results are contained in Table~\ref{Tab-A-M}. The difference in
estimated mean scores, which is negative in $5/12$ cases under
consideration, is not significantly $\neq0$ each time.

%
\begin{table}
\caption{List of share name abbreviations used in Table \protect\ref
{Tab-A-M}}\label{Tab-A-abbr}
\begin{tabular*}{\textwidth}{@{\extracolsep{\fill}}lcccc@{}}
\hline
\textbf{Abbreviation} & \textbf{Full name} & & \textbf{Abbreviation} &
\textbf{Full name} \\
\hline
ADS.DE & Adidas & & FRE.DE & Fresenius VZ \\
ALV.DE & Allianz & & HEI.DE & Heidelbergcement \\
BEI.DE & Beiersdorf & & HEN3.DE & Henkel VZ \\
BMW.DE & BMW & & MRK.DE & Merck \\
DAI.DE & Daimler & & MUV2.DE & Munich RE \\
DBK.DE & Deutsche Bank & & RWE.DE & RWE \\
EOAN.DE & E.ON & & SIE.DE & Siemens \\
FME.DE & Fresenius Medical Care & & & \\
\hline
\end{tabular*}
\end{table}

%
\begin{table}[b]
\tabcolsep=0pt
\caption{Mean scores for univariate and multivariate VaR estimation
($\alpha= 0.01$) for the log-returns of several stocks (date values
starting from at least 2003-01-01); for full names of shares,
cf. Table \protect\ref{Tab-A-abbr}}
\label{Tab-A-M}
{\fontsize{8.7}{10.7}\selectfont
\begin{tabular*}{\textwidth}{@{\extracolsep{\fill}}lccccd{2.2}d{2.2}cd{2.2}c@{}}
\hline
 &  &  & \multicolumn
{2}{c}{\textbf{Mean scores}} & \multicolumn{1}{c}{} &
\multicolumn{1}{c}{} & & \\[-6pt]
 &  &  & \multicolumn
{2}{c}{\hrulefill} &  & & & \\
\multicolumn{1}{c}{\multirow{2}{45pt}[10pt]{\textbf{Share N}$\bolds{^{\circ}}$\textbf{1}
\textbf{Abbr.}}} &
 \multicolumn{1}{c}{\multirow{2}{43pt}[10pt]{\centering\textbf{Share N}$\bolds{^{\circ}}$~\textbf{2} \textbf{Abbr.}}}
&\multicolumn{1}{c}{\textbf{Corr.}} &
\multicolumn{1}{c}{$\bolds{\hat m_{N, \F}}$} &
\multicolumn{1}{c}{$\bolds{\hat
m_{N, \cG}}$} &
\multicolumn{1}{c}{\multirow{2}{56pt}[10pt]{\centering \textbf{Diff.}
$\bolds{(=M_N)}$ $\bolds{\hat m_{N, \F} - \hat m_{N, \cG
}}$}} & \multicolumn{1}{c}{\multirow{2}{42pt}[10pt]{\centering\textbf{Rel. diff.}
$\bolds{M_N/\hat m_{N, \F}}$}} & \multicolumn
{1}{c}{$\bolds{\hat\sigma}$} & \multicolumn{1}{c}{$\bolds{T_N}$} & \multicolumn
{1}{c}{$\bolds{\operatorname{Pr}(>\!T_N)}$} \\
\hline
FME.DE & HEI.DE & 0.104 & 4.80 &
4.76 & 0.04 & 0.01 & 7.1 & 0.30 & 0.383 \\
ADS.DE & FME.DE & 0.186 & 3.98 & 4.00 & -0.02 & -0.01 & 3.3 & -0.34 &
0.632 \\
ADS.DE & BEI.DE & 0.204 & 4.02 & 4.02 & 0.00 & 0.00 & 3.1 & 0.05 &
0.480 \\
FME.DE & MRK.DE & 0.218 & 4.51 & 4.50 & 0.00 & 0.00 & 3.7 & 0.02 &
0.492 \\
FRE.DE & HEN3.DE & 0.227 & 3.91 & 3.84 & 0.07 & 0.02 & 3.2 & 1.04 &
0.150 \\
FME.DE & HEN3.DE & 0.266 & 3.84 & 3.87 & -0.03 & -0.01 & 2.6 & -0.51 &
0.694 \\[3pt]
DAI.DE & SIE.DE & 0.654 & 5.78 & 5.90 & -0.12 & -0.02 & 3.5 &
-1.76 & 0.961 \\
EOAN.DE & RWE.DE & 0.680 & 6.24 & 6.25 & -0.01 & -0.00 & 6.7 & -0.04 &
0.517 \\
BMW.DE & DAI.DE & 0.719 & 5.54 & 5.56 & -0.02 & -0.00 & 2.4 & -0.40 &
0.654 \\
ALV.DE & DBK.DE & 0.744 & 6.01 & 6.00 & 0.01 & 0.00 & 1.8 & 0.25 &
0.401 \\
ALV.DE & MUV2.DE & 0.766 & 5.24 & 5.22 & 0.03 & 0.00 & 1.8 & 0.72 &
0.237 \\
\hline
\end{tabular*}}
\end{table}

\textit{Conclusions}.
Using the models under consideration, there seems to be small potential
for improvement by using the multivariate DCC-model for the individual
risk factors instead of the simple $\operatorname{GARCH}(1,1)$-model for the portfolio
returns. However, further investigations with distinct multivariate
time series models would be required.

%

\section{Concluding remarks}\label{sec:conclusions}

Additional information should lead to better forecasts, at least if the
forecasting mechanism is ideal, that is based on the true conditional
distribution.
But how can the improvement of an increase in information on the
forecast, for example, the mean, a quantile or the whole predictive
distribution, be quantified, what exactly is improved?

The answer that we give in this paper is in terms of the expected loss
(score) under a strictly consistent scoring function or rule, which is
attuned to the predicted parameter. This interpretation is particularly
attractive if the expected loss is by itself of interest. For instance,
for the value at risk (a quantile), we show that the expected loss
under an appropriate scoring function turns out to be the expected shortfall.

While for ideal forecasts, additional information is thus always useful
or at least not harmful, this is apparently no longer true if
information (data) needs to be processed by a statistician
in terms of model building, selection and estimation before making
predictions. For example, in our application on value at risk
prediction for log-returns, it turned out that a multivariate modeling
of individual risk factors often performs worse than a simple
univariate modeling of the portfolio returns.

Thus, the development of model selection criteria with the aim of
optimal prediction of a certain parameter under a specific scoring
function, such as the AIC for the mean and squared error in regression
models, should be a major issue of future research.

\begin{appendix}\label{app}
\section*{Appendix}
\subsection*{Proofs}
We start with the following well-known fact, which we prove for lack of
reference.
\begin{lemma}\label{lemma:measprobmeas}
Let $(\Omega, \F, P)$ be a probability space, and let $ G_\F\dvtx \Omega
\times\B\to[0, 1]$ be a Markov kernel for which $G_\F(\omega,
\cdot) \in\Theta$ for all $\omega\in\Omega$. Then $G_\F\dvtx \Omega
\to\Theta$, $\omega\mapsto G_\F(\omega, \cdot)$ is $\F- \B
(\Theta)$ measurable.
\end{lemma}
\begin{pf}
For a fixed continuous, bounded function $f\dvtx D \to\R$, the map
%
\begin{equation}
\label{eq:meashelp} \omega\mapsto\int_\R f(x)
G_{\F}(\omega;dx)
\end{equation}
is $\F-\mathcal{B}$-measurable; see \citeauthor{MR2372119}
[(\citeyear{MR2372119}), Theorem~8.37].
The weak topology on $\Theta$ may be metrized by
\[
d(\mu, \nu) = \sup_n \biggl\{\biggl |\int f_n \,d \mu-
\int f_n \,d \nu \biggr| \dvtx\mu, \nu\in\Theta \biggr\},
\]
where $(f_n)$ is an appropriate sequence of bounded, continuous
functions on $D$; see \citeauthor{MR1385671} (\citeyear{MR1385671}),
Theorem~1.12.2.

The metric space $(\Theta, d)$ is separable; see
\citeauthor{MR2372119} (\citeyear{MR2372119}),
page~252. Therefore, for the measurability of $G_\F$, it
suffices to show that the preimage of every closed ball $B_\varepsilon
(\mu)$, $\varepsilon>0$, $\mu\in\Theta$ in the metric $d$, under
$G_\F$ is in $\F$. Now,
\[
B_{n,\varepsilon}(\mu) = \biggl\{ \omega\in\Omega\dvtx \biggl|\int f_n(x)
G_\F(\omega;dx) - \int f_n(x) \,d \mu(x) \biggr| \leq\varepsilon
\biggr\} \in\F
\]
by (\ref{eq:meashelp}) and, hence, also
\[
G_\F^{-1} \bigl(B_\varepsilon(\mu) \bigr) = \bigcap
_n B_{n,\varepsilon
}(\mu) \in\F.
\]
\upqed\end{pf}

\begin{pf*}{Proof of Theorem~\ref{the:basic}}
Let $\mu_{Y | \F}$ denote the conditional distribution with
corresponding conditional distribution functions $F_{Y | \F}$.
Since by the above lemma the map $\mu_{Y | \F} \dvtx \Omega\to\Theta$,
$\omega\mapsto\mu_{Y | \F}(\omega, \cdot)$, is $\F-\mathcal
{B}(\Theta)$-measurable, and since by assumption $T$ is $\mathcal
{B}(\Theta) - \mathcal B$-measurable, it follows that $\hat Y (\omega
)= T \circ\mu_{Y|\F}(\omega; \cdot)$ is an
$\F$-measurable random variable.

For $P\mbox{-a.e. }  \omega\in\Omega$,
\[
E \bigl(S(Z,Y)|\F \bigr) (\omega) = \int_\R S \bigl(Z(
\omega), y \bigr) F_{Y|\F}(\omega, dy).
\]
Since $S$ is strictly consistent, for all $\omega\in\Omega$ we have
\[
\int_\R S \bigl(\hat Y(\omega), y \bigr)
F_{Y|\F}(\omega, dy) \leq\int_\R S \bigl(Z(
\omega), y \bigr) F_{Y|\F}(\omega, dy)
\]
with equality if and only if $Z(\omega) = \hat Y(\omega)$. The second
statement follows by taking expectated values. 
\end{pf*}

\begin{pf*}{Proof of Corollary~\ref{cor:condineq}}
The proof of the first statement of (\ref{eq:ineqsecondincrease}) is
immediate from Theorem~\ref{the:basic}, since $\hat Y_\F$ is also
$\cG$-measurable.
For the second, take conditional expectation w.r.t.~$\F$. Since for a
nonnegative random variable $Z$, $Z = 0$ a.s. if and only if $E(Z |\F
) =0$ a.s., the second conclusion follows. For the third, take
unconditional expectation.
\end{pf*}

\begin{pf*}{Proof of Theorem~\ref{the:characpropscorerule}}
Set $X(\omega) = \bS (G_\F(\omega,\cdot), Y(\omega)  )$.
By Lemma~\ref{lemma:measprobmeas}, $X$ is measurable. Then for $P\mbox{-a.e. }
 \omega\in\Omega$,
\[
E (X|\F ) (\omega) = \int_\R\bS \bigl(G_\F(
\omega, \cdot ), y \bigr) F_{Y|\F}(\omega, dy).
\]
Since $\bS$ is strictly proper, for all $\omega\in\Omega$ we have
\[
\int_\R\bS \bigl(F_{Y|\F}(\omega, \cdot), y
\bigr) F_{Y|\F
}(\omega, dy) \leq\int_\R\bS
\bigl(G_\F(\omega, \cdot), y \bigr) F_{Y|\F
}(\omega, dy)
\]
with equality if and only if the distributions $F_{Y|\F}(\omega,
\cdot)$ and $G_\F(\omega, \cdot)$ coincide. This proves the first
part of the theorem, the second follows by taking unconditional
expected values.
The final statement is a standard fact of probability.
\end{pf*}

\begin{pf*}{Proof of Theorem~\ref{the:testsuffinfo}}
Set
\[
W_n = \frac{1}{n} \sum_{k=1}^n
\bigl(S\bigl(\hat Y^{(h)}_{k, \F}, Y_k\bigr) - S
\bigl(\hat Y^{(h)}_{k, \cG}, Y_k\bigr) \bigr) =
\frac{1}{n} \sum_{k=1}^n
Z_k.
\]
Under an alternative, Corollary~\ref{cor:condineq}, (\ref{eq:ineqsecondincrease}),
 $3$rd statement, implies that $E
Z_1 >0$, and the ergodic theorem then implies $\sqrt{n} W_n \to\infty
$, $P$-a.s.

From (\ref{eq:asympnoeffect}) with $O_P(\sqrt{n})$, $ \sqrt{n}
(M_n - W_n) = O_P(1)$, therefore, $\sqrt{n} M_n \to\infty$, $P$-a.s.
as well.

Under the null hypothesis, from Corollary~\ref{cor:condineq}, (\ref
{eq:ineqsecondincrease}), $1$st statement with equality
implies that $E(Z_n|\cG_{n-h}) = 0$ for all $n$. Therefore, setting $\|
X\|_2 = (E X^2)^{1/2}$, we have that
\[
\sum_{n=0}^\infty\bigl\|E(Z_0 |
\cG_{-n})\bigr\|_2 = \sum_{n=0}^{h-1}
\bigl\|E(Z_0 | \cG_{-n})\bigr\|_2 < \infty,
\]
and from the CLT for stationary sequences [see \citet{Durrett2005PTAE},
Theorem~7.6,
page~416]
\[
\sqrt{n} W_n \stackrel{d} {\to} N \bigl(0, \sigma^2
\bigr),
\]
where $\sigma^2$ is as in (\ref{eq:asymptest}). From (\ref
{eq:asympnoeffect}) with $o_P(\sqrt{n})$, $ \sqrt{n}  (M_n - W_n) =
o_P(1)$, therefore, asymptotic normality holds true for $\sqrt{n} M_n$
as well.
\end{pf*}

\begin{pf*}{Proof of Proposition~\ref{prop:independ}}
We only show the implication $1 \Rightarrow2$. By independence, we
have that for $P\mbox{-a.e. } \omega\in\Omega$,
\[
F_{Y|\F} \bigl(\omega, \bigl(Z(\omega), \infty\bigr) \bigr) = P(I=1|\F )
(\omega) = P(I=1) = 1-\alpha,
\]
so that $Z(\omega) = q_{\alpha} (F_{Y|\F}(\omega, \cdot) )$.
\end{pf*}
\begin{pf*}{Proof of Corollary~\ref{cor:indexceed}}
This follows from the fact that the $(I_n)$ are independent if and only
if for all $n$, $I_{n+1}$ and $\sigma(I_k; k \leq n)$ are independent.
\end{pf*}

\begin{pf*}{Proof of Proposition~\ref{prop:meansscorequant}}
For a strictly increasing, continuous distribution function $F$, a
simple calculation gives that
\[
E_F \bigl(S^*\bigl(q_\alpha(F), Y\bigr) \bigr) = -
\frac{1}{\alpha} \int_{-\infty}^{q_\alpha(F)} y \,dF(y).
\]
Therefore, (\ref{eq:increaseinfoquant}) follows from (\ref
{eq:ineqsecondincrease}).
\end{pf*}
\end{appendix}

\section*{Acknowledgments}
The authors would like to thank the Editor Tilmann Gneiting as well as
three anonymous referees for helpful comments and for pointing out
several relevant references, and Steffen Dereich for helpful discussions.
%

%



\printaddresses

\end{document}